\documentstyle[epsf,twocolumn,seceq]{jpsj}

\def\runtitle{Heat Capacity of the Pressure-Induced
Superconductivity in itinerant Ferromagnet UGe$_2$}
\def\runauthor{Naoyuki {\sc TATEIWA} {\it et al}}

\title
{
Heat Capacity of the Pressure-Induced Superconductivity in
Itinerant Ferromagnet UGe$_2$
}

\author
{ 
Naoyuki {\sc
TATEIWA}\footnote{E-mail:tateiwa@djebel.mp.es.osaka-u.ac.jp}
, Tatsuo C KOBAYASHI$^{1}$, Kiichi AMAYA, Yoshinori
HAGA$^{2}$, Rikio SETTAI$^{3}$ and Yoshichika \=ONUKI$^{3}$
}

\inst
{
Department of Physical Science, Grasuate School of
Engineering Science, Osaka University, Toyonaka, Osaka
560-8531, Japan\\
$^1$Research Center for Materials Science at Extreme
Conditions, Osaka University, Toyonaka, Osaka 560-8531,
Japan\\
$^2$ Advanced Science Research Center, Japan Atomic Energy
Research Institute, Tokai-mura, Naka-Gun, Ibaraki, 319-1195,
Japan\\
$^3$ Department of Physics, Graduate School of Science,
Osaka University, Toyonaka, Osaka 560-0043, Japan\\

}

\recdate
{
\today
}

\abst
{
Recently co-existence of the ferromagnetism and
superconductivity was reported in the high-pressure region
(1.0-1.6 GPa) in UGe$_2$. We performed the heat capacity
measurement on UGe$_2$ under high pressure. At 1.13 GPa, we
found a peak corresponding to the superconducting
transition. The superconducting temperature $T_{\rm SC}$ and
${\it\Delta} C/(\gamma T_{\rm SC})$ are 0.6 K and 0.25, 
respectively. The superconducting transition was also
confirmed by the appearance of the Meissner effect in the
{\it ac} susceptibility. From these results, we confirm a
bulk nature of the superconductivity in UGe$_2$. The value
of $C/T$ ($\sim$ 95 mJ/moleK$^2$) just above $T_{\rm SC}$ at
1.15 GPa is as much as 3 times larger than that at ambient
pressure, which indicates a large mass enhancement of
quasiparticles under high pressure.
}

\kword
{
UGe$_2$, Ferromagnetism, Superconductivity
}

\begin{document}
\sloppy
\maketitle
\section{Introduction}
   UGe$_2$ is a metallic ferromagnet with a Curie
temperature $T_{\rm C}$ = 52 K and the spontaneous moment
1.4 $\mu_{\rm B}$/U at ambient pressure~\cite{rf:1,rf:2}.
The magnetization is highly anisotropic along the easy {\it
a}-axis in orthorhombic crystal structure. Recently it was
reported that UGe$_2$ is a pressure-induced
superconductor~\cite{rf:3,rf:4,rf:5}. Superconductivity was
observed in the pressure range from 1.0 to 1.6 GPa.
Surprisingly, in this pressure range UGe$_2$ is still in the
ferromagnetic state. We have preformed the heat capacity
measurement on UGe$_2$ in order to confirm the bulk nature
of the superconductivity. 
\section{Experimental}
  A single crystal was grown by the Czochralski pulling
method in a tetra-arc furnace as described in ref. 5. The
residual resistivity ratio RRR of the present sample were
600 at ambient pressure. The heat capacity and the {\it ac}
susceptibility were measured by the adiabatic heat pulse and
Hartshorn bridge methods, respectively. The low-temperature
range of 0.18-6K was generated by $^3$He-$^4$He dilution
refrigerator. Pressure was applied by utilizing a Cu-Be
piston-cylinder cell with a Daphne oil (7373) as a
pressure-transnitting medium.
\section{Result and Discussion}
  Figure 1 shows temperature dependence of $C/T$ for $P$ = 0
and 1.15 GPa, together with the {\it ac} susceptiblity. 
\begin{figure}
 \begin{center}
  \epsfxsize=9cm
  \epsfbox{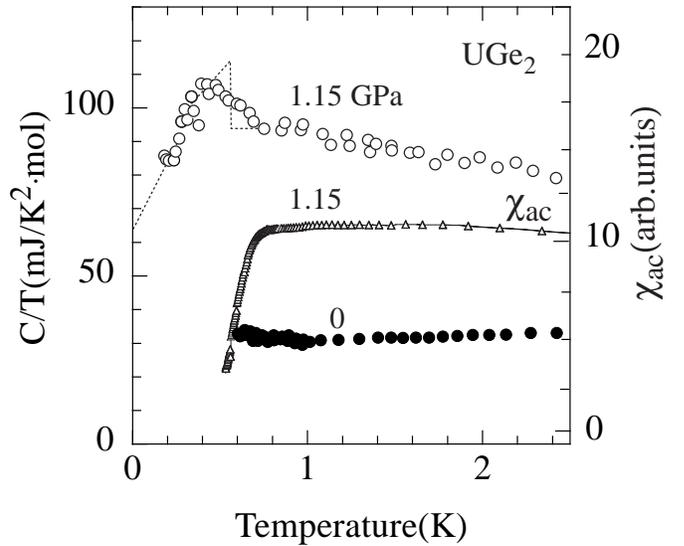}
 \end{center}
\caption{Temperature dependence of the heat capacity in the
form of $C/T$ at 0 and 1.15 GPa, together with the $ac$
susceptiblity at 1.15 GPa in UGe$_2$. }
\label{fig:1}
\end{figure} 

\begin{figure}
 \begin{center}
  \epsfxsize=8.5cm
  \epsfbox{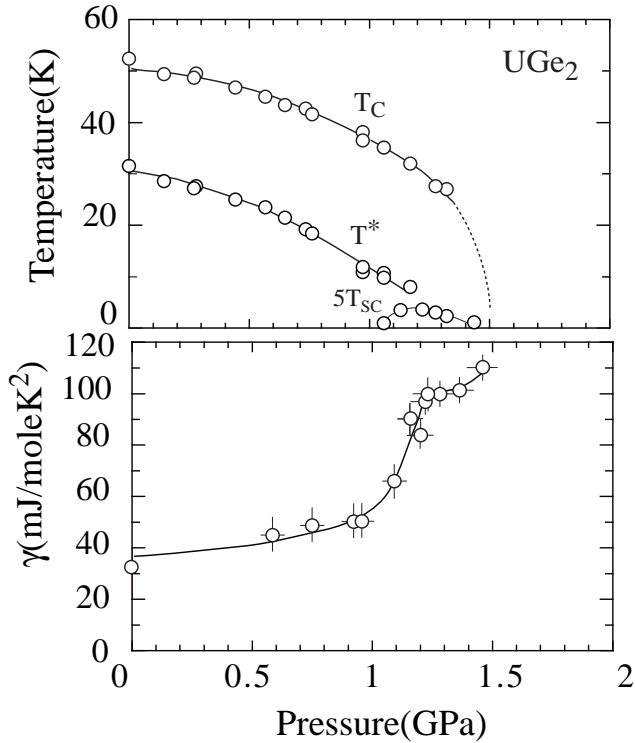}
 \end{center}
\caption{Pressure phase diagram and pressure dependence of
the $\gamma$ value in UGe$_2$.}
\label{fig:1}
\end{figure}  
 The {\it ac} susceptibility was measured for the same
sample, simultaneously with the heat capacity measurement.
At ambient pressure, $C/T$ is almost constant at low
temperature and the $\gamma$ value is determined to be about
30 mJ/moleK$^2$, which is consistent with the previous
report~\cite{rf:1}. 
At 1.15 GPa, we found a peak in $C/T$ around 0.6 K
corresponding to the superconducting transition. The
superconducting transition temperature $T_{\rm SC}$ is about
0.6 K. The superconducting transition was also confirmed by
the appearance of the Meissner effect in the {\it ac}
susceptibility below $T_{\rm SC}$. From these results, it
was concluded that the superconductivity in UGe$_2$ was a
bulk phenomenon.

  The peak of $C/T$ at $T_{\rm SC}$ is weak as ${\it\Delta}
C/(\gamma T_{\rm SC})$ = 0.25, which is smaller than a BCS
value of 1.45. Here ${\it\Delta} C$ is the jump of the heat
capacity at $T_{\rm SC}$. The residual $\gamma$ value of 65
mJ/moleK$^2$ is large. Machida {\it et al.} discussed the
superconductivity of UGe$_2$ on the basis of a non-unitary
triplet paring~\cite{rf:6}. In this mechanism, only the
Fermi surface of the majority band opens the gap and the
minority band remains normal state below $T_{\rm SC}$. 
Recent band calculations suggest that there is substantial
contibution to the density of state at the Fermi energy from
the minority band~\cite{rf:7,rf:8}. Thus present large
residual $\gamma$ value might be explained by this scenario.
However we suggest that the large residual $\gamma$ is a
consequence of the sample quality because the
superconducting peak is broad. Considering that the present
sample is a high quality single crystal sample with RRR=
600, the superconductivity in UGe$_2$ seems to be very
sensitive to small amount impurities. It suggests the
unconventional nature of the superconductivity and the
pairing symmetry is not s-wave but possibly p-wave.
\begin{figure}
 \begin{center}
  \epsfxsize=8.5cm
  \epsfbox{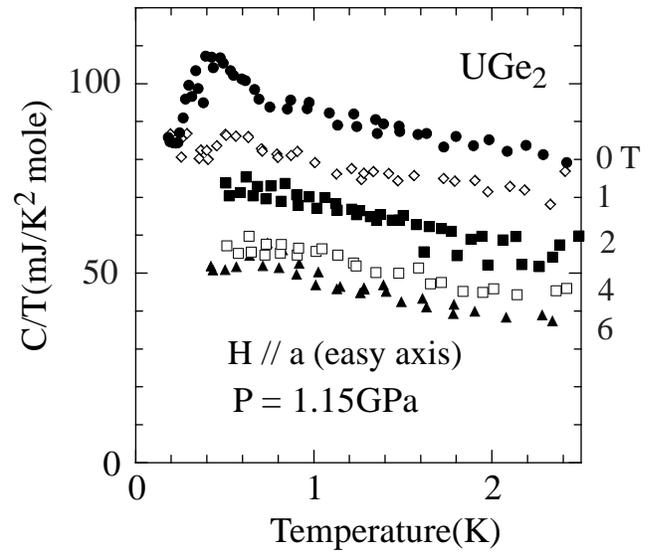}
 \end{center}
\caption{Temperature dependence of the heat capacity in the
form of $C/T$ at1.15 GPa  under the magnetic field up to 6 T
in UGe$_2$.}
\label{fig:1}
\end{figure} 

  The $\gamma$ value at 1.15 GPa was determined to be 95
mJ/moleK$^2$ from the value of $ C/T$ just above $T_{\rm
SC}$.  This $\gamma$ value is about three times larger than
that at ambient pressure. Figure 2 shows the pressure
dependence of the $\gamma$ value, together with the phase
diagram in UGe$_2$. $T_{\rm C}$ is the Curie temperature.
$T^{*}$ is another characteristic temperature where the
resistivity and the magnetization show an
anomaly~\cite{rf:4,rf:9}. The detailed nature of the
transition at $T^{*}$ is not known at present. The $\gamma$
value increases steeply above 1.0 GPa, where
superconductivity starts to appear, and has a shoulder at
1.2 GPa where $T^{*}$ becomes zero and $T_{\rm SC}$  shows
maximum. This suggests that a large $\gamma$ value
originated from the low energy magnetic fluctuation related
to the phase transition at $T^{*}$ and this magnetic
fluctuation might be driving force for the formation of
Cooper pairs.

   Figure 3 shows the temperature  dependence of $C/T$ at
1.15 GPa under the magnetic field up to 6 T along the ${\it
a}$ axis (easy axis). The superconducting peak was not
observed in the mangetic field of 1 T. $C/T$ was strongly
depressed by the appplication of magnetic fields. It
suggests that a low energy magnetic fluctuiation, which is
the origin of the large $\gamma$, is suppressed by the
magnetic field. 

   In conclusion, we confirm the bulk superconductivity in
UGe$_2$ by the heat capacity measurement. The large $\gamma$
value observed in the pressure region where
superconductivity appears suggests the superconductivity of
heavy quasiparticles.

{\bf Ackonowledgement}

   This work was financially supported by the Grant-in Aid
for COE Research (No. 10CE2004) of the Ministry of
Education, Science, Sports and Culture, and by CREST, Japan
Science and Technology Corporation.

\end{document}